\documentclass[reprint, aps, pre, amsmath,amssymb,groupedaddress,superscriptaddress]{revtex4-2}

\usepackage{graphicx}
\usepackage{dcolumn}
\usepackage{bm}
\usepackage{color}
\usepackage{enumitem}
\usepackage[normalem]{ulem}
\usepackage{comment}

\newif\ifHighlitedChanges
\def\ifHighlitedChanges{\iftrue}
\def\ifHighlitedChanges{\iffalse}
\ifHighlitedChanges
  
  \def\STRIKE#1{{\color{red}\sout{#1}}}
\else
  
  \def\STRIKE#1{\relax}
\fi

\begin{document}
\title{Energy transport between heat baths with oscillating temperatures}
\author{Renai Chen}
\affiliation{Theoretical Division and Center for Nonlinear Studies, Los Alamos National Laboratory, Los Alamos, New Mexico, USA}
\author{Tammie Gibson}
\affiliation{Theoretical Division, Los Alamos National Laboratory, Los Alamos, New Mexico, USA}
\author{Galen T. Craven}
\email[]{galen.craven@gmail.com}
\affiliation{Theoretical Division, Los Alamos National Laboratory, Los Alamos, New Mexico, USA}
\begin{abstract}
Energy transport is a fundamental physical process that plays a prominent role in the function and performance of myriad systems and technologies.
Recent experimental measurements have shown that subjecting a macroscale system to a time-periodic temperature gradient can increase thermal conductivity in comparison to a static temperature gradient.
Here, we theoretically examine this mechanism in a nanoscale model by applying
a stochastic Langevin framework to describe the energy transport properties of a particle connecting two heat baths with different temperatures, where the temperature difference between baths is oscillating in time.
Analytical expressions for the energy flux of each heat bath and for the system itself 
are derived for the case of a free particle and a particle in a harmonic potential.
We find that 
dynamical effects in the energy flux induced by temperature oscillations give rise to complex energy transport hysteresis effects.
The presented results suggest that applying time-periodic temperature modulations 
is a potential route to control energy storage and release in molecular devices and nanosystems.
\end{abstract}

\maketitle

\section{Introduction}
Discovering and understanding 
the fundamental physical mechanisms governing energy transport processes at the nanoscale is one of the most important problems
in the molecular sciences \cite{Cahill2002, Cahill2003, Dhar2008,Dubi2011,Sato2012,Maldovan2013,Segal2016,Ness2016,Ness2017,Nascimento2022}.
Vibrational, electronic, and radiative energy transport mechanisms, as well as their interplay,
manifest at the nanoscale in complex nonequilibrium processes
 \cite{Li2012, Segal2016,Sabhapandit2012,Lebowitz1959,Lebowitz1967,Lebowitz1971,Lebowitz2008,Lebowitz2012,Nitzan2003thermal,Segal2005prl,Lebowitz2012,Dhar2015,Velizhanin2015,Esposito2016,craven16c,matyushov16c,craven17a,craven17b,craven17e,craven20a,Ochoa2022}.
Gaining a deeper understanding of these mechanisms is becoming increasingly important in order to advance the development of multiple technologies.
Energy transport 
plays a prominent role in the function of physical, biological, and technological systems \cite{Leitner2008,Dubi2011,Li2012}.
Therefore, developing accurate theoretical tools to describe energy transport processes is critical.
At the nanoscale, controlling energy transport in the form thermal energy, i.e., heat, has broad applications to advance the design of
electronic devices \cite{Cahill2003,Nitzan2007,Ratner2013review,Lim2013}, 
thermoelectric molecules and materials \cite{Esposito2015,Russ2016organic,Cui2017perspective,Gehring2021}, 
and phononic systems that use heat to perform logical operations \cite{Li2012,Donadio2015,Velizhanin2015}. 
Recent advances in the ability to probe energy transport at the atomistic level, both
theoretically \cite{chen2020local,sharony2020stochastic} and experimentally \cite{cui2019thermal,mosso2019thermal},
facilitate the use and manipulation of energy transport properties in practical molecular applications.

There have been significant recent advances in the experimental set-ups used to examine nanoscale heat transport.
Specifically, several groups have developed experimental techniques capable of measuring thermal conductance at the single-molecule level \cite{cui2019thermal,mosso2019thermal}. 
These advances open the possibility to probe, characterize, and utilize energy transport effects by tailoring molecular structural characteristics.  
Energy transport through  molecular structures typically involves complex nonequilibrium dynamics \cite{Nitzan2003thermal,Donadio2015, craven15a, craven15c,craven17d, craven17c} due to the interplay between multiple heat transport mechanisms at the nanoscale.
Advanced theoretical tools are therefore needed to describe these nanoscale processes, which typically do not follow 
macroscale principles, for example, Fourier's law \cite{BonettoFourier2000,Bonetto2004SCR,Segal2009SCR,Chang2008,craven2023a}. 


At the macroscale, periodic modulation of a temperature gradient can enhance heat flow 
leading to increased thermal conductivity in comparison to a static temperature gradient \cite{urban2022thermal}.
However, the molecular origins of this enhanced thermal conductivity have not been established. 
Here, we explore how  the thermal transport properties of a model nanoscale system 
can be modified by applying periodic temperature modulations. 
Time-dependent temperatures can be used 
to affect, control, and probe molecular properties of systems ranging in size from large macromolecules to single atoms \cite{goss1998quantitative,platkov2014periodic,singleatom2016}.
Temperature modulations play a prominent role in the function of a multitude of systems including: 
pyroelectric materials \cite{Bowen2014pyro,Yamamoto2021,Lheritier2022pyro},
thermal batteries \cite{Gur2012thermalbattery,Wang2022battery},
molecular ratchets \cite{bartussek1994periodically,Hanggi2009wireperiodictemps,Zhang2008ratchet},
thermal devices with memory  \cite{Ben-Abdallah2017thermalmemristor,Ordonez-Miranda2019thermalmemristor},
and calorimetry devices \cite{Seid2011calorimetry,Shoifet2015calorimetry}.

In this article, we use a paradigmatic model of a nanoscale system to examine the energy transport properties of a particle connecting two heat baths with different temperatures, where the temperature difference between baths is periodic in time (See Fig.~\ref{fig:TOC}).
Theoretical frameworks describing the thermodynamics of a system in contact with one or more heat baths with time-dependent temperatures have a rich history \cite{Reimann2002,brey1990generalized,hern07a,hern13d,Ford2015,Seifert2015periodictemp,Seifert2016periodiccurrent,Awasthi2021,Portugal2022effective,Volz2022,Weron2022}.
However, these previously developed theoretical formalisms cannot be used to accurately describe the model we consider in this article because of one or more of the following: (a) they are constructed for the case of a system in contact with a single heat bath and therefore do not describe energy transport properties between multiple baths with different temperatures,
(b) they are derived from macroscale principles that do not accurately capture nanoscale energy transport properties,
and/or 
(c) they assume the system is in a quasistatic state and therefore do not describe the limit of fast temperature oscillations.

The specific theoretical formalism we apply is a stochastic Langevin equation describing a single particle that bridges two heat baths with oscillating temperatures.
The time-dependent temperature difference between the two baths affects the energy transport through the system (the particle) and results in phenomena, for example, energy flux hysteresis, that are not present in the case of a static temperature difference.
Analytical expressions for the different energy fluxes in the model, specifically the energy flux from each bath and from particle itself, 
are derived for the case of a free particle and a particle in a harmonic potential.
The results of stochastic molecular dynamics simulations are in strong agreement with the derived results,
supporting the validity of the derived energy flux expressions. 
We find that the energy transport properties can be significantly altered by the temperature oscillations.

The remainder of the article is organized as follows: 
Section \ref{sec:model} contains a description of the model used to examine energy transport between heat baths with oscillating temperatures.  
In Sec.~\ref{sec:HT}, the definitions and general formalism for energy transport are presented,
including definitions of how the energy fluxes in the model are determined.
Section \ref{sec:flux} contains derivations of the energy flux expressions for
two cases: (a) a free particle (Sec.~\ref{subsec:free}) and (b) a particle in a harmonic potential (Sec.~\ref{subsec:harmonic}).
Results and discussion about each case are included in the corresponding section. 
Conclusions and future directions are presented in Sec.~\ref{sec:conclusion}.

\section{Model}
\label{sec:model}

\begin{figure}[]
\includegraphics[width = 8.5cm,clip]{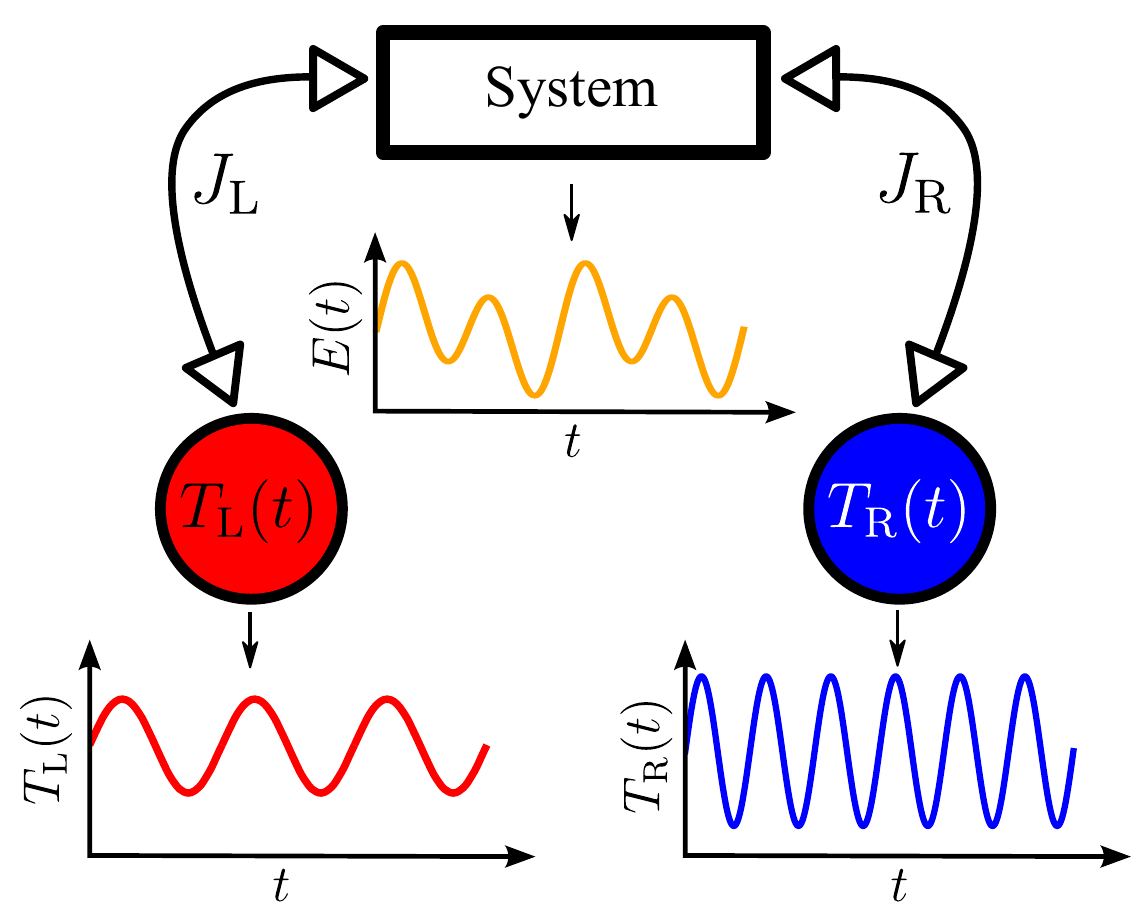}
\caption{\label{fig:TOC}
Schematic diagram of the model used in this work. 
The system, represented by a labeled rectangle, is in contact with two heat baths, $\text{L}$(left) and  $\text{R}$(right). Bath $\text{L}$ is represented by a red circle and bath $\text{R}$ is represented by a blue circle.
The time-dependent temperatures of the baths are $T_\text{L}(t)$ and $T_\text{R}(t)$.
The temperatures of each bath are oscillating in time, as illustrated by the graphs at the bottom of the figure.
The energy of the system $E(t)$ oscillates due to the temperature oscillations.
The black curves with arrows illustrate the energy flow channels in the model.
The energy fluxes in/out of each bath are $J_\text{L}$ and $J_\text{R}$.
}
\end{figure}

The model we use 
to examine nanoscale heat transport in the presence of temperature oscillations 
consists of a single particle that is in contact with two heat baths, both of which 
have temperatures that are oscillating in time.
Specifically, the two heat baths, denoted by $\text{L}$ for ``left'' bath and $\text{R}$ for ``right'' bath, have periodically oscillating temperatures
$T_\text{L}(t) = T_\text{L}(t+ \mathcal{T}_\text{L})$ and $T_\text{R}(t) = T_\text{R}(t+ \mathcal{T}_\text{R})$ 
where $\mathcal{T}_\text{L}$ and $\mathcal{T}_\text{R}$ are the respective periods of oscillation.
A schematic diagram of the model is shown in Fig.~\ref{fig:TOC}.
The Langevin equation of motion for the system is
\begin{equation}
\begin{aligned}
\label{eq:EoM1}
\dot x &= v, \\
\dot v &= - \gamma_\text{L} \dot x  - \gamma_\text{R} \dot x - m^{-1}\partial_x U(x) +  \xi_\text{L}(t) + \xi_\text{R}(t), 
\end{aligned}
\end{equation}
where $x$ is the position of the particle, $v$ is the particle velocity, $U(x)$ is the potential energy, 
$\gamma_\text{L}$ and $\gamma_\text{L}$ are dissipative (friction) terms for each
bath that parameterize the system-bath coupling strength, and $\xi_\text{L}(t)$ and $\xi_\text{R}(t)$ are stochastic noise terms that obey
the following correlations:
\begin{equation}
\begin{aligned}
\label{eq:noise}
 \big\langle \xi_\text{L}(t) \xi_\text{L}(t')\big\rangle &= 2 \gamma_\text{L} k_\text{B} m^{-1} T_\text{L}(t)\delta(t-t'), \\[0ex]
 \big\langle \xi_\text{R}(t) \xi_\text{R}(t')\big\rangle &= 2 \gamma_\text{R} k_\text{B} m^{-1} T_\text{R}(t)\delta(t-t'), \\[0ex]
	\big\langle \xi_\text{L}(t) \xi_\text{R}(t')\big\rangle &= 0,  \\[0ex]
 \big\langle \xi_\text{L}(t)\big\rangle &=0, \\[0ex]
 \big\langle \xi_\text{R}(t)\big\rangle &=0, \\[0ex]
\end{aligned}
\end{equation}
where $m$ is the particle mass and $k_\text{B}$ is the Boltzmann constant.
The notation $\langle \ldots \rangle$ denotes an average over realizations of the noise.
These correlations are at the Markovian limits describing the physical situation in which the intensity (strength) of the noise of each bath is oscillating in time, and that intensity depends on the temperature of that bath at the current time $t$ but not on the temperature at previous times $t'$. 
Theoretical formulations of time-dependent temperatures that include non-Markovian effects have been developed \cite{hern07a,hern13d}.
Here we will only examine the Markovian case.
Specifically, Eq.~(\ref{eq:EoM1}) is a special two-bath memoryless limit of the generalized Langevin equation derived in Ref.~\citenum{hern07a}.
We will examine two cases for the potential: a free particle with $U(x) =0$ and a particle in a harmonic potential $U(x) = \tfrac{1}{2}mkx^2$.
We define the temperatures of each bath to take the specific oscillatory forms
\begin{align}
 T_\text{L}(t)   &= T^{(0)}_\text{L} + \Delta T_\text{L} \sin(\omega_\text{L} t),\\[1ex]
 T_\text{R}(t)   &= T^{(0)}_\text{R} + \Delta T_\text{R} \sin(\omega_\text{R} t),
\end{align}
where $T^{(0)}_\text{L}$ and $T^{(0)}_\text{R}$ are the temperatures of the two baths in the limit of vanishing of oscillations, $\Delta T_\text{L}$ and $\Delta T_\text{R}$ define the amplitude of the oscillations, and $\omega_\text{L}$ and $\omega_\text{R}$ are oscillation frequencies. 
In the case in which $\omega_\text{L}$ and $\omega_\text{R}$ are commensurable, 
the system and heat currents are periodic, and we denote the total period of the model as $\mathcal{T}$.
The instantaneous temperature difference between the two baths is 
\begin{equation}
\Delta T(t) =    T_\text{L}(t) - T_\text{R}(t).
\end{equation}

\section{Heat Transport}
\label{sec:HT}
The energetic properties of the system can be described using the Sekimoto formalism for stochastic energetics \cite{Sekimoto1998}
which involves
separating the terms in Eq.~(\ref{eq:EoM1}) into contributions to the system energy change and contributions to the energy change in the baths. 
We specifically separate the expected energy fluxes in our model into  three terms:
\begin{enumerate}
  \item $J_\text{sys}$ is the energy flux in/out of the system. It is generated by changes in the energy of the system. Part of the energy transferred by this flux will be partitioned to the right bath and part will be partitioned to the left bath. 
  \item $J_\text{L}$ is the energy flux associated with the left bath.
  \item $J_\text{R}$ is the energy flux associated with the right bath.
\end{enumerate}
The sum of the energy fluxes obeys conservation of energy:
$J_\text{L} (t)+J_\text{R} (t) + J_\text{sys} (t) = 0$.
Using the stochastic energetics formalism, the expected energy fluxes in/out of the baths and the system (see Fig.~\ref{fig:TOC}) can be expressed as \cite{Lebowitz1959,Sekimoto1998,Sabhapandit2012,Dhar2015}:
\begin{align}
\label{eq:heatcurrentbathL}
J_\text{L} (t) &=  m \gamma_\text{L} \big\langle  v^2(t) \big\rangle-m \big\langle \xi_\text{L}(t) v(t)\big\rangle ,\\[1ex]
\label{eq:heatcurrentbathR}
J_\text{R} (t) &=  m \gamma_\text{R} \big\langle  v^2(t) \big\rangle-m \big\langle \xi_\text{R}(t) v(t)\big\rangle ,\\[1ex]
\label{eq:heatcurrentbathS}
J_\text{sys} (t) &= \partial_t \big\langle E(t)\big\rangle,
\end{align}
where $E(t)$ is the energy of the system, $\big\langle  v^2(t) \big\rangle$ is the second velocity moment of the system, and $\big\langle \xi_\text{L}(t) v(t)\big\rangle$ and $\big\langle \xi_\text{R}(t) v(t)\big\rangle$ are noise-velocity correlation functions.
We use a sign convention such that
the energy flux expressions are positive when energy enters the corresponding bath/system
and negative when energy leaves the bath/system.
The system energy flux can be separated into the sum of two parts 
\begin{equation}
J_\text{sys} (t) = J^{(\text{L})}_\text{sys} (t) +  J^{(\text{R})}_\text{sys} (t),
\end{equation}
where $J^{(\text{L})}_\text{sys}$ and $J^{(\text{R})}_\text{sys}$ are the part of the system energy flux that flows in/out of the left bath and right bath, 
respectively.

Because the bath temperatures are varying, 
the system will not reach a nonequilibrium steady state (NESS). 
Instead, the system approaches a time-dependent nonequilibrium state (TDNS)
with an average energy that is oscillating in time.
In a system that is in a NESS, $\partial_t \big\langle E(t)\big\rangle = 0$ and $J_\text{L} (t) = -J_\text{R} (t)$.
However, in the case of periodic temperatures the system energy can be a time-dependent quantity leading to a nonvanishing derivative of the energy with respect to time.

Our analysis will focus on examining the heat current properties over a period of oscillation.
The expected heat that is obtained/released by the baths or the system $\text{K}\in \{\text{L}, \text{R}, \text{sys}\}$ over the time interval  $[0,\mathcal{T}]$  is
\begin{align}
\label{eq:heat}
 \mathcal{Q}_\text{K} &=  \int_0^{\mathcal{T}} J_\text{K} (t') dt'.
\end{align}
For the purpose of examining energy storage capabilities of the system, it will often be useful to separate the total energy change $\mathcal{Q}_\text{K}$ into two parts \cite{Fu2020},
the energy obtained by the bath/system
\begin{equation}
\label{eq:heatplus}
\mathcal{Q}^+_\text{K} =  \int_0^{\mathcal{T}} J_\text{K} (t') \Theta[J_\text{K} (t') ]  dt',
\end{equation}
and the energy lost by the bath/system
\begin{equation}
\label{eq:heatminus}
\mathcal{Q}^-_\text{K} =  \int_0^{\mathcal{T}} J_\text{K} (t') \Theta[-J_\text{K} (t') ]  dt',
\end{equation}
where $\Theta$ is the Heaviside function.

\section{Energy Flux Derivation and Results}
\label{sec:flux}

\subsection{Free Particle}
\label{subsec:free}
We first examine the case of free particle by setting $U(x) = 0$. 
The equation of motion for a free particle
connected to two thermal baths can be expressed as 
\begin{equation}
\begin{aligned}
\label{eq:EoMBrownian}
\dot x &= v, \\
\dot v &= - \gamma \dot x  + \xi_\text{L}(t) + \xi_\text{R}(t), 
\end{aligned}
\end{equation}
with 
\begin{equation}
\gamma = \gamma_\text{L}+\gamma_\text{R},
\end{equation}
being the effective friction.
It is important to note that in our analysis we will not assume that the system is in a quasistatic nonequilibrium state.
Because we do not make the quasistatic assumptions, inertial effects due to the oscillating temperatures can  significantly affect the system's heat transport properties.

The equation of motion~(\ref{eq:EoMBrownian}) is solved by the set of equations:
\begin{equation}
\begin{aligned}
x(t) &= x_0 + \int_0^t v(s) \,ds, \\[1ex]
\label{eq:vsol}
v(t) &=  v_0 e^{- \gamma t}  \\
&\quad + \int_0^t  e^{-\gamma(t-s)}\xi_\text{L}(s)\,ds + \int_0^t  e^{-\gamma(t-s)}\xi_\text{R}(s)\,ds,
\end{aligned}
\end{equation}
where $x_0$ is the initial position and $v_0$ is the initial velocity of the particle.
These formal solutions can be applied to construct expressions for the moments and time-correlation functions of a 
particle driven by two thermal sources.
The average energy of the system, that is of the particle, is
\begin{align}
\label{eq:energyBrownian}\big\langle E(t)\big\rangle  &= \frac{1}{2} m \big\langle v^2(t)\big\rangle.
\end{align}
Note that in the {\it absence} of temperature oscillations, i.e., in the $(\Delta T_\text{L}, \Delta T_\text{R}) \to (0,0)$ limit, the system reaches a nonequilibrium steady state 
in which the energy of the system is \cite{Sekimoto1998, Zamponi2005, hern07a, craven18a1, craven18a2}
\begin{equation}
\langle E\rangle  = \frac{1}{2}k_\text{B} T,
\end{equation}
where 
\begin{equation}
\label{eq:temp}
T = \frac{\gamma_\text{L} T^{(0)}_\text{L} + \gamma_\text{R} T^{(0)}_\text{R}}{\gamma_\text{L}+\gamma_\text{R}},
\end{equation}
is the effective temperature of the system.

In order to evaluate the energy flux for a free particle, we will need to evaluate the correlation functions in Eqs.~(\ref{eq:heatcurrentbathL})-(\ref{eq:heatcurrentbathS}).
The noise-velocity correlation functions $\left\langle \xi_\text{L}(t) v(t)\right\rangle$ and $\left\langle \xi_\text{R}(t) v(t)\right\rangle$ 
for a free particle can be constructed
using Eq.~(\ref{eq:vsol}),
\begin{align}
\label{eq:velnoisecorrL}
\nonumber\big\langle \xi_\text{L}(t) v(t)\big\rangle &= 
\big\langle \xi_\text{L}(t) v_0 \big\rangle  e^{-\gamma t}
+\int_0^t e^{-\gamma (t-s)} \big\langle\xi_\text{L}(t)\xi_\text{L}(s)\big\rangle \,ds \\
\nonumber &\quad +\int_0^t e^{-\gamma (t-s)} \big\langle\xi_\text{L}(t)\xi_\text{R}(s)\big\rangle \,ds \\
& = \frac{\gamma_\text{L} k_\text{B} T_\text{L}(t)}{m},\\
\label{eq:velnoisecorr}
\nonumber\big\langle \xi_\text{R}(t) v(t)\big\rangle &= 
\big\langle \xi_\text{R}(t) v_0 \big\rangle  e^{-\gamma t}
+\int_0^t e^{-\gamma (t-s)} \big\langle\xi_\text{R}(t)\xi_\text{L}(s)\big\rangle \,ds \\
\nonumber &\quad +\int_0^t e^{-\gamma (t-s)} \big\langle\xi_\text{R}(t)\xi_\text{R}(s)\big\rangle \,ds \\
& = \frac{\gamma_\text{R} k_\text{B} T_\text{R}(t)}{m},
\end{align} 
where we have utilized $\big\langle \xi_\text{L}(t) v_0 \big\rangle = \big\langle \xi_\text{R}(t) v_0 \big\rangle  = 0$ 
and the correlations in Eq.~(\ref{eq:noise})
to complete the evaluation. 

The other correlation function that must be evaluated in order to obtain analytical expressions for the energy fluxes is $\left\langle v^2(t) \right\rangle$.
Squaring the formal solution in Eq.~(\ref{eq:vsol}), and applying the correlations in Eq.~(\ref{eq:noise}) yields:
\begin{widetext}
\begin{equation}
\begin{aligned}
\label{eq:vsqrBrownian}
\big\langle v^2(t)\big\rangle  &=  
 v^2_0 e^{- 2\gamma t}
 + \int_0^t\!\!\int_0^t  e^{-\gamma(2t-s_1-s_2)}\big\langle\xi_\text{L}(s_1)\xi_\text{L}(s_2)\big\rangle\,ds_1\,ds_2 
+ \int_0^t\!\!\int_0^t  e^{-\gamma(2t-s_1-s_2)}\big\langle\xi_\text{R}(s_1)\xi_\text{R}(s_2)\big\rangle\,ds_1\,ds_2\\[1ex]
& \quad + 2\!\int_0^t e^{-\gamma(2 t-s_1)}\big\langle  \xi_\text{L}(s_1)v_0\big\rangle\,ds_1
+2 \!\int_0^t e^{-\gamma(2t-s_1)}\big\langle  \xi_\text{R}(s_1)v_0 \big\rangle\,ds_1 
 + 2\!\int_0^t\!\!\int_0^t  e^{-\gamma(2t-s_1-s_2)}\big\langle\xi_\text{L}(s_1)\xi_\text{R}(s_2)\big\rangle\,ds_1\,ds_2,\\[1ex]
&= v^2_0 e^{-2 \gamma t}
+ \frac{k_\text{B}T}{m} \left(1-e^{-2 \gamma t}\right) \\[1ex]
&  
\quad + \frac{2 k_\text{B}}{m}\bigg(\frac{ \gamma_\text{L} \Delta T_\text{L} (2 \gamma \sin(\omega_\text{L} t) - \omega_\text{L} \cos(\omega_\text{L} t)+ \omega_\text{L} e^{-2 \gamma t})}{4 \gamma^2+\omega^2_\text{L}} 
+ \frac{\gamma_\text{R} \Delta T_\text{R} (2 \gamma \sin(\omega_\text{R} t) - \omega_\text{R} \cos(\omega_\text{R} t)+\omega_\text{R} e^{-2 \gamma t})}{4 \gamma^2+\omega^2_\text{R} }\bigg).
\end{aligned}
\end{equation}
\end{widetext}
The last three terms in the top equation vanish because $\big\langle \xi_\text{L}(t) v_0 \big\rangle = 0$, $\big\langle \xi_\text{R}(t) v_0 \big\rangle = 0$, and $\big\langle \xi_\text{L}(t) \xi_\text{R}(t') \big\rangle = 0$.
In the limit of vanishing temperature oscillations, $\big\langle v^2(t) \big\rangle = v^2_0 e^{-2 \gamma t} + k_\text{B} T / m \left(1-e^{-2 \gamma t}\right)$ where the effective temperature $T$ is defined in Eq.~(\ref{eq:temp}).
In the long-time limit this expression reduces to $\big\langle v^2(t) \big\rangle =  k_\text{B} T / m$.
A fraction of the total energy flux is 
energy that is obtained/released by the particle,
the rest being heat current between baths.

In the long time limit, the system approaches a time-dependent (oscillatory) nonequilibrium state. 
In this limit, the exponential terms vanish
and the energy flux expressions simplify to
\begin{widetext}
\begin{align}
\label{eq:JLlong}
J_\text{L}(t) &= 
\gamma_\text{L} k_\text{B} \bigg( T - T_\text{L} (t) 
+ \frac{2\gamma_\text{L} \Delta T_\text{L} (2 \gamma \sin(\omega_\text{L} t) - \omega_\text{L} \cos(\omega_\text{L} t))}{4 \gamma^2+\omega^2_\text{L}} 
+ \frac{2\gamma_\text{R} \Delta T_\text{R} (2 \gamma \sin(\omega_\text{R} t) - \omega_\text{R} \cos(\omega_\text{R} t))}{4 \gamma^2+\omega^2_\text{R} }
 \bigg), \\[1ex]
\label{eq:JRlong}
J_\text{R}(t) &=  
\gamma_\text{R} k_\text{B} \bigg( T -T_\text{R} (t) + \frac{2\gamma_\text{L} \Delta T_\text{L} (2 \gamma \sin(\omega_\text{L} t) - \omega_\text{L} \cos(\omega_\text{L} t))}{4 \gamma^2+\omega^2_\text{L}} 
+ \frac{2 \gamma_\text{R} \Delta T_\text{R} (2 \gamma \sin(\omega_\text{R} t) - \omega_\text{R} \cos(\omega_\text{R} t))}{4 \gamma^2+\omega^2_\text{R} }
 \bigg),\\[1ex]
 \label{eq:JSlong}
 J_\text{sys}(t) &=  k_\text{B}\bigg( \frac{  \gamma_\text{L} \Delta T_\text{L} \omega_\text{L} (2 \gamma \cos(\omega_\text{L} t) + \omega_\text{L} \sin(\omega_\text{L} t))}{4 \gamma^2+\omega^2_\text{L}} 
+ 
\frac{  \gamma_\text{R} \Delta T_\text{R} \omega_\text{R}  (2 \gamma \cos(\omega_\text{R} t) + \omega_\text{R} \sin(\omega_\text{R} t))}{4 \gamma^2+\omega^2_\text{R}}
 \bigg), 
\end{align}
\end{widetext}
which are independent of the initial velocity $v_0$.
Equations~(\ref{eq:JLlong})-(\ref{eq:JSlong}) are primary results of this subsection. 
They are analytical expressions for the energy fluxes generated in the TDNS.
It is also important to note that because the system is in a TDNS, 
the net energy flux from the system $J_\text{sys}(t)$ does not vanish.
It can be verified that by combining Eqs.~(\ref{eq:JLlong}) - (\ref{eq:JSlong}), the conservation of energy relation
\begin{equation}
J_\text{L} (t)+J_\text{R} (t) + J_\text{sys} (t) = 0
\end{equation}
is satisfied.  

In the limit of quasistatic $(\text{QS})$  temperature oscillations, at every time instant the system is characterized by a Gibbs distribution with effective temperature 
\begin{equation}
\label{eq:tempQS}
T^{(\text{QS})}(t) = \frac{\gamma_\text{L} T_\text{L}(t) + \gamma_\text{R} T_\text{R}(t)}{\gamma_\text{L}+\gamma_\text{R}}.
\end{equation}
The QS limit
implies that the temperature of the bath is varying slowly enough such that the relaxation rate of the system $\gamma$ is much faster than the oscillation rates of both temperatures.
In the QS limit defined by $(\omega_\text{L}/\gamma, \omega_\text{R}/\gamma) \to (0,0)$,
the energy flux expressions reduce to

\begin{align}
\label{eq:JLlongsteadystate}
J^{(\text{QS})}_\text{L}(t) &=  k_\text{B} \frac{\gamma_\text{L} \gamma_\text{R}}{\gamma_\text{L} + \gamma_\text{R}} \Big(T_\text{R}(t)-T_\text{L}(t)\Big),\nonumber\\[1ex]
J^{(\text{QS})}_\text{R}(t) &= k_\text{B} \frac{\gamma_\text{L} \gamma_\text{R}}{\gamma_\text{L} + \gamma_\text{R}} \Big(T_\text{L}(t)-T_\text{R}(t)\Big), \\[1ex]
J^{(\text{QS})}_\text{sys}(t) &= 0.\nonumber
\end{align}

In the limit of vanishing temperature oscillations $(\Delta T_\text{L}, \Delta T_\text{R}) \to (0,0)$, which we will refer to as the static $(\text{S})$ limit, the energy flux expressions reduce to the forms 
\begin{align}
\label{eq:JLlongsteadystate_notime}
J^{(\text{S})}_\text{L} &=  k_\text{B} \frac{\gamma_\text{L} \gamma_\text{R}}{\gamma_\text{L} + \gamma_\text{R}} \Big(T^{(0)}_\text{R}-T^{(0)}_\text{L}\Big),\nonumber\\[1ex]
J^{(\text{S})}_\text{R} &= k_\text{B} \frac{\gamma_\text{L} \gamma_\text{R}}{\gamma_\text{L} + \gamma_\text{R}} \Big(T^{(0)}_\text{L}-T^{(0)}_\text{R}\Big),\\[1ex]
J^{(\text{S})}_\text{sys} &= 0, \nonumber
\end{align}
which do not depend on time.
These are well-known forms for the heat current of a single particle in contact with two heat baths with different temperatures \cite{Lebowitz1959}.

\begin{figure}[ht]
\includegraphics[width = 8.5cm,clip]{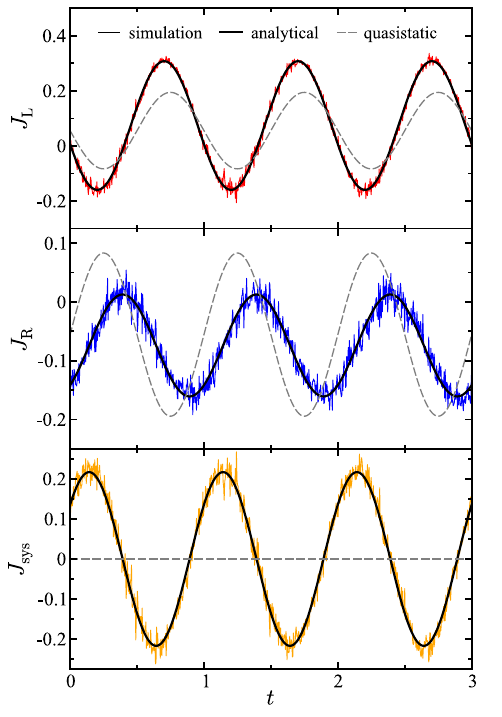}
\caption{\label{fig:Flux_1}
Time-dependence of the energy fluxes for the left bath (top), right bath (middle), and the system (bottom) 
in the case of a free particle connecting two heat baths with oscillating temperatures.
Each energy flux is shown in units of $\gamma k_\text{B} T$.
Time is shown in units of the total oscillation period $\mathcal{T}$.
Parameters are $\gamma = 2$ ($\gamma_\text{L} = 1$, $\gamma_\text{R} = 1$), $m = 1$, $T^{(0)}_\text{L} = 0.8$, $T^{(0)}_\text{R} = 1.07$,
$\Delta T_\text{L} = 0.5$, $\Delta T_\text{R} = 0$, $\omega_\text{L} = 5$, $\omega_\text{R} = 0$.
All parameters throughout are given in reduced units with 
characteristic dimensions: $\widetilde{\sigma} = 1\,\text{\AA}$,  $\widetilde{\tau} = 1\,\text{ps}$,
$\widetilde{m} = 10\,m_u$,
and $\widetilde{T} = 300\,\text{K}$.
In each panel, the black curve is the exact analytical result, the dashed gray curve is the result given by the corresponding quasistatic expression, and the noisy colored curves are the results generated from molecular dynamics simulations.
}
\end{figure}

Shown in Fig.~\ref{fig:Flux_1} are the energy fluxes of the left bath, the right bath, and the system
for the situation in which the temperature of one bath is oscillating and the temperature of the other bath is constant. 
Both the analytical results and the results of  molecular dynamics (MD) simulations are shown. 
The MD results are generated by integrating Eq.~(\ref{eq:EoM1}) using the Euler-Maruyama scheme and then calcualting the energy fluxes using the stochastic energetics formalism \cite{Sabhapandit2012,Sekimoto1998}. 
The parameters for all simulations in this article are given in reduced units with 
characteristic dimensions: $\widetilde{\sigma} = 1\,\text{\AA}$,  $\widetilde{\tau} = 1\,\text{ps}$,
$\widetilde{m} = 10\,m_u$,
and $\widetilde{T} = 300\,\text{K}$.
The results of the MD simulations
are in excellent agreement with the analytical results
for all three energy fluxes in the model.
It can be observed that
even though the temperature of the right bath is not oscillating,
the effect of the oscillating left bath temperature propagates through the system,
causing  periodic fluctuations in the energy flux of the right bath.
In the quasistatic limit,
the energy fluxes of the left and right baths
have the same magnitude but opposite signs
while the system energy flux is zero at all times.
Significant differences in the energy flux oscillation phase and magnitude 
are observed between the exact analytical result and the quasistatic result.
These effects will play an important role 
in the generation of the energy flux hysteresis, which is discussed later.

\begin{figure}[t]
\includegraphics[width = 8.5cm,clip]{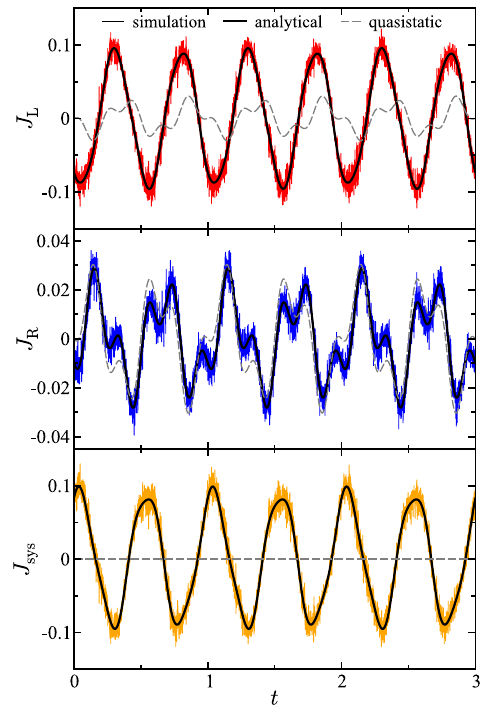}
\caption{\label{fig:Flux_2}
Time-dependence of the energy fluxes for the left bath (top), right bath (middle), and the system (bottom) 
in the case of a free particle connecting two heat baths with oscillating temperatures.
Each energy flux is shown in units of $\gamma k_\text{B} T$.
Time is shown in units of the total oscillation period $\mathcal{T}$.
Parameters are $\gamma = 1.7$ ($\gamma_\text{L} = 1.5$, $\gamma_\text{R} = 0.2$), $m = 1$, $T^{(0)}_\text{L} = 1$, $T^{(0)}_\text{R} = 1$,
$\Delta T_\text{L} = 0.2$, $\Delta T_\text{R} = 0.1$, $\omega_\text{L} =2$, $\omega_\text{R} = 5$.
In each panel, the black curve is the exact analytical result, the dashed gray curve is the result given by the corresponding quasistatic expression, and the noisy colored curves are the results generated from molecular dynamics simulations.
}
\end{figure}

The energy fluxes for the case in which the temperatures of 
both baths are oscillating at different frequencies are shown in Fig.~\ref{fig:Flux_2}. 
The system-bath couplings are asymmetric, with the coupling being stronger for the left bath than for the right bath, $\gamma_\text{L}>\gamma_\text{R}$.
In this case, complex dynamics are observed. 
The oscillation frequency of the strongly-coupled left bath is dominate in the energy flux of the left bath
while the weakly-coupled right bath exhibits features of multiple frequencies, meaning it is influenced by the oscillations of both baths. 
The system energy flux exhibits a complex pattern, combining the functional characteristics 
of the left bath and right bath energy fluxes.
This because all the energy that is transported 
from the left bath to right bath goes through system,
hence the system takes on characteristics of both bath energy fluxes.
In the quasistatic limit,
the left and right fluxes 
are a mix of the two sinusoidal temperature oscillation patterns,
and, interestingly, are more aligned with the weakly-coupled right bath oscillation pattern.
The system energy flux vanishes in the quasistatic limit.
The MD results are in excellent agreement the analytical results over all these complex trends.

\begin{figure}[]
\includegraphics[width = 8.5cm,clip]{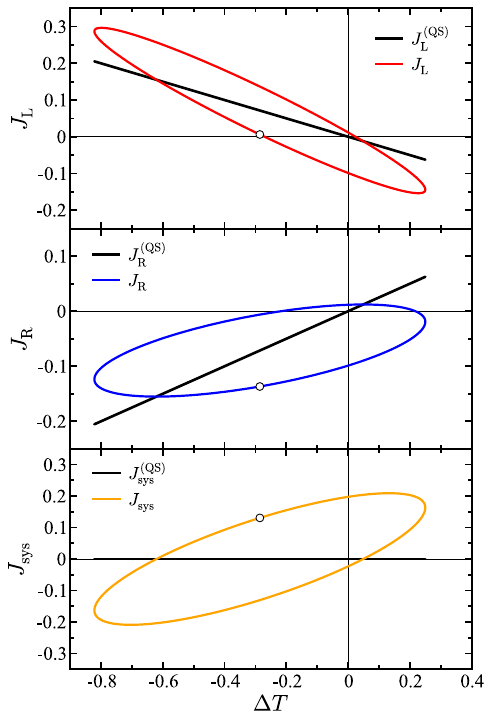}
\caption{\label{fig:Hys_1}
Energy flux as a function of temperature difference for the left bath (top), right bath (middle), and the system (bottom) which is a free particle.
The parameters are the same as Fig.~\ref{fig:Flux_1}.
Each energy flux is shown in units of $\gamma k_\text{B} T$ and the temperature difference $\Delta T$ is shown in units of $T$.
In each panel, the colored curve is the exact analytical result and the solid black line is the result given by the quasistatic expression.
The marker on each curve marks the value of the respective energy flux at $t=0$.
}
\end{figure}

Hysteresis effects in electronic fluxes are of significant interest in the field of neuromorphic computing and, more broadly, in the development of logical devices with memory \cite{Caravelli2018,Van2018organic,Sangwan2020neuromorphic,Yuriy2011}.
Dynamical effects in the energy fluxes induced by fast temperature oscillations relative to the system relaxation rate (the system-bath coupling) can generate hysteresis effects.
We will refer to these effects as intertial effects.
Using the derived energy flux expressions,
we find that in the limit of fast temperature oscillations in comparison to the system relaxation timescale $\gamma$,
the energy transport properties can be significantly different than in the quasistatic limit of slow temperature oscillations.
In Fig.~\ref{fig:Hys_1}, the energy fluxes in the model are shown as a function of the time-dependent temperature difference between baths $\Delta T(t)$. 
The $t=0$ starting point in each hysteresis loop is denoted by a circular marker.
Using the exact analytical results, 
hysteresis is observed in all of the energy fluxes,
meaning that the same temperature difference can generate different flux values
depending on the time in the temperature oscillation period.
This differs from the quasistatic limit where the hysteresis effects vanish,
as shown by the solid black lines.
We also observe transient energy transport direction 
that goes against the thermal gradient, that is, time periods in the oscillation cycle where heat flows in the cold to hot direction.
However, this effect is only transient, and the net heat flow, i.e., the average energy flux, is always in the 
direction (hot to cold) that obeys thermodynamics principles.
The quasistatic results are shown as black lines in each panel. In this limit,
no hysteresis effects are observed.

\begin{figure}[]
\includegraphics[width = 8.5cm,clip]{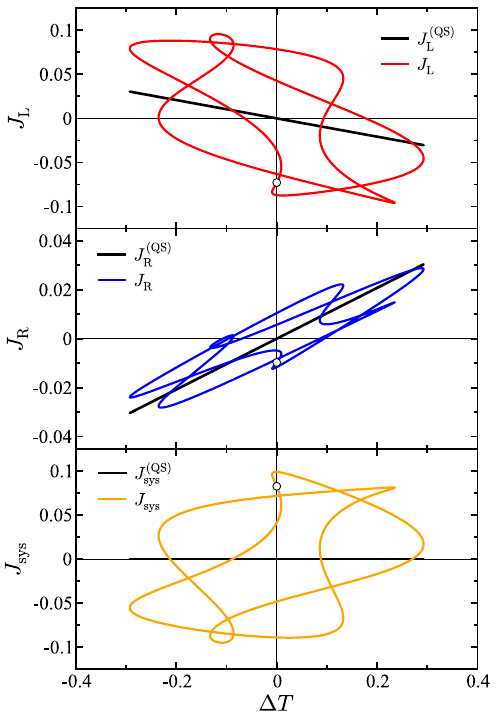}
\caption{\label{fig:Hys_2}
Energy flux as a function of temperature difference for the left bath (top), right bath (middle), and the system (bottom) which is a free particle.
The parameters are the same as Fig.~\ref{fig:Flux_2}.
Each energy flux is shown in units of $\gamma k_\text{B} T$ and the temperature difference $\Delta T$ is shown in units of $T$.
In each panel, the colored curve is the exact analytical result and the solid black line is the result given by the quasistatic expression.
The marker on each curve marks the value of the respective energy flux at $t=0$.
}
\end{figure}

Energy flux hysteresis loops are shown in Fig.~\ref{fig:Hys_2} 
for the case in which the temperatures of both baths are oscillating, but at different frequencies.
Because of inertial effects brought on by fast temperature oscillations, the energy fluxes are 
not zero when $\Delta T = 0$.
The patterns generated are reminiscent of Lissajous curves,
but the interweaving paths and asymmetric shapes result in complex loops 
and should be attributed to the complexities that arise from fast temperature 
oscillations and the resulting TDNS energy transport processes.
Therefore the energy flux paths are different from the typical Lissajous shapes.
The relationships between the temperature bias and the energy flux that give rise to hysteresis are only possible when the temperature oscillations are fast relative to the system relaxation rates
and so the system is not in a quasistatic state.

\begin{figure}[]
\includegraphics[width = 8.5cm,clip]{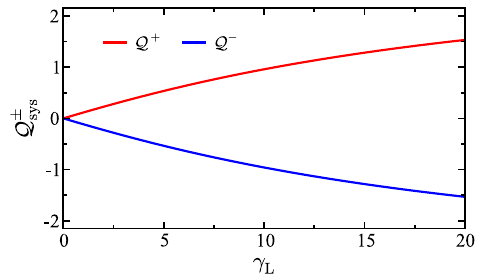}
\caption{\label{fig:Qpm_free}
Energy obtained (red top curve) and released (blue bottom curve) by the system  over one period of temperature oscillation  as function of the coupling parameter $\gamma_\text{L}$ with 
$\gamma_\text{R} = 2$ held constant.
All other parameters are the same as Fig.~\ref{fig:Flux_1}.
Energy is shown in units of $ k_\text{B} T$.
The coupling parameter is shown in units of $1/\mathcal{T}$.
}
\end{figure}

As described by the equations in Section~\ref{sec:HT},
the energy fluxes will accumulate and dissipate over a certain period of time.
The magnitude of energy storage and energy release are insightful to calculate,
as they relate how much total energy is being stored and released by the system.
This is analogous to electric charges accumulating in electric circuits.
In Fig.~\ref{fig:Qpm_free} the energy obtained by the system over a period of oscillation
is shown as a function of variation of the left system-bath coupling $\gamma_\text{L}$ while holding the right coupling $\gamma_\text{R}$ constant.
The red line is the energy obtained by the system
during one period of temperature oscillation
and the blue line is energy released from the system during the same period.
As the coupling strength to the left bath is increased,
the energy obtained increases monotonically.
The energy obtained and the energy released have equal magnitude but opposite signs.
This illustrates that over a period of oscillation there is no net energy storage in the system.
More complicated patterns and trends arise when extra layers of complexity are added,
e.g., when non-zero potential energy forms are used as described in the next section (See also Fig.~\ref{fig:Qplus_har}).

\subsection{Harmonic Potential}
\label{subsec:harmonic}

Next, we examine the case in which the particle connecting the two heat baths moves in a harmonic potential. The equation of motion for a particle in the potential $U(x) = \tfrac{1}{2} m k x^2$ can be written as
\begin{equation}
\begin{aligned}
\label{eq:EoMHar}
\dot x &= v, \\
\dot v &= - \gamma \dot x -k x  + \xi_\text{L}(t) + \xi_\text{R}(t), 
\end{aligned}
\end{equation}
with $\gamma = \gamma_\text{L}+\gamma_\text{R}$ as before.
For convenience we write Eq.~\ref{eq:EoMHar} as
\begin{equation}
\label{eq:eomnonhar}
		\begin{pmatrix}
			\dot{x}(t)\\
			\dot{v}(t) 
		\end{pmatrix}=
\begin{pmatrix}
			0&1 \\
			-k &-\gamma
		\end{pmatrix}
		\begin{pmatrix}
			x(t)\\
			v(t)
		\end{pmatrix}
		+
		\begin{pmatrix}
			0\\
			\xi_\text{L}(t) +\xi_\text{R}(t)
		\end{pmatrix}.
\end{equation}
The complementary equation of Eq.~(\ref{eq:eomnonhar}) is
\begin{equation}
\label{eq:eomhomohar}
		\begin{pmatrix}
			\dot{x}_\text{c}(t)\\
			\dot{v}_\text{c}(t) 
		\end{pmatrix}=
		\mathbf{A}
		\begin{pmatrix}
			x(t)\\
			v(t)
		\end{pmatrix}.
\end{equation}
with 
\begin{equation}
		\mathbf{A}=
		\begin{pmatrix}
			0&1 \\
			-k &-\gamma
		\end{pmatrix}. \\ 
\end{equation}
Eigenvalues of $\mathbf{A}$ are
\begin{equation}
\begin{aligned}
\lambda_1 &= \frac{1}{2}\left(-\gamma - \sqrt{\gamma^2-4 k}\right),\\
\lambda_2 &= \frac{1}{2}\left(-\gamma + \sqrt{\gamma^2-4 k}\right).
\end{aligned}
\end{equation}
For simplicity, we define
\begin{equation}
\Delta \lambda = \lambda_2 - \lambda_1.
\end{equation}
The fundamental matrix solution of (\ref{eq:eomhomohar}) is
\begin{equation}
\mathbf{M}(t) = 	\left(\mathbf{v}_1 e^{\lambda_1 t} \;\; \mathbf{v}_2 e^{\lambda_2 t} \right),
\end{equation}
where $\mathbf{v}_1$ and $\mathbf{v}_2$ are eigenvectors.
The solution of the initial value problem is 
\begin{equation}
		\begin{pmatrix}
			x_\text{c}(t)\\
			v_\text{c}(t)
		\end{pmatrix} = \mathbf{M}(t)\mathbf{M}^{-1}(0)\begin{pmatrix}
			x_0\\
			v_0
		\end{pmatrix},
\end{equation}
which can expressed as
\begin{equation}
\begin{pmatrix}
		x_\text{c}(t)\\
			v_\text{c}(t) 
		\end{pmatrix}	
		=  
		 \begin{pmatrix}
			\dfrac{\displaystyle e^{\lambda_1 t} \left(\lambda_2 x_0 -v_0 \right)-  e^{\lambda_2 t}\left( \lambda_1 x_0 -v_0 \right)}{\displaystyle \Delta \lambda}\\[1.5ex]
		\dfrac{\displaystyle \lambda_1 e^{\lambda_1 t}\left(\lambda_2 x_0 -v_0 \right)-\lambda_2 e^{\lambda_2 t}\left(\lambda_1 x_0 -v_0 \right) }{\displaystyle \Delta \lambda}
		\end{pmatrix}.
\end{equation}
The critical damping $\lambda_1 = \lambda_2$ solution is obtained by taking the $\lambda_1 \to \lambda_2$ limit in the previous equations.

The formal solution of the stochastic equation of motion (\ref{eq:eomnonhar}) is
\begin{equation}
\begin{aligned}
\label{eq:formal}
\begin{pmatrix}
			x(t)\\
			v(t) 
		\end{pmatrix} 
		&= \mathbf{M}(t)\mathbf{M}^{-1}(0)
		\begin{pmatrix}
			x_0\\
			v_0
		\end{pmatrix}\\
		&\quad +\int_0^t \mathbf{M}(t)\mathbf{M}^{-1}(s) \begin{pmatrix}
			0\\
			\xi_\text{L}(s) +\xi_\text{R}(s)
		\end{pmatrix}\,ds,
\end{aligned}		
\end{equation}
where the first term on the RHS is the complementary part of the solution.
The formal solution can be written as
\begin{align}
\label{eq:xsolhar}
\nonumber &x(t) = \frac{\displaystyle  e^{\lambda_1 t} \left(\lambda_2 x_0 -v_0 \right)-  e^{\lambda_2 t}\left(\lambda_1 x_0 -v_0 \right)}{\displaystyle \Delta \lambda} \\
\nonumber & - \frac{1}{\Delta \lambda}\left(\int_0^t  e^{\lambda_1(t-s)}\xi_\text{L}(s)\,ds 
+ \int_0^t  e^{\lambda_1(t-s)}\xi_\text{R}(s)\,ds\right)\\
& + \frac{1}{\Delta \lambda}\left(\int_0^t  e^{\lambda_2(t-s)}\xi_\text{L}(s)\,ds 
+ \int_0^t  e^{\lambda_2(t-s)}\xi_\text{R}(s)\,ds\right) ,
\\[1ex]
\label{eq:vsolhar}
\nonumber&v(t) =  \frac{\displaystyle \lambda_1 e^{\lambda_1 t}\left(\lambda_2 x_0 -v_0 \right)-\lambda_2 e^{\lambda_2 t}\left(\lambda_1 x_0 -v_0 \right) }{\displaystyle \Delta \lambda}  \\
\nonumber&- \frac{\lambda_1}{\Delta \lambda}\left(\int_0^t  e^{\lambda_1(t-s)}\xi_\text{L}(s)\,ds 
+ \int_0^t  e^{\lambda_1(t-s)}\xi_\text{R}(s)\,ds\right)\\
& + \frac{\lambda_2}{\Delta \lambda}\left(\int_0^t  e^{\lambda_2(t-s)}\xi_\text{L}(s)\,ds 
+ \int_0^t  e^{\lambda_2(t-s)}\xi_\text{R}(s)\,ds\right).
\end{align}

Using the formal solutions (\ref{eq:xsolhar}) and (\ref{eq:vsolhar}) and the noise correlations in Eq.~(\ref{eq:noise}), 
the noise-velocity correlation functions $\big\langle \xi_\text{L}(t) v(t)\big\rangle$ and $\big\langle \xi_\text{R}(t) v(t)\big\rangle$ in the heat current expressions in Eqs.~(\ref{eq:heatcurrentbathL}) and (\ref{eq:heatcurrentbathR}) can be written as 
\begin{widetext}
\begin{align}
\label{eq:velnoisecorrLhar}
\nonumber\big\langle \xi_\text{L}(t) v(t)\big\rangle &= 
\big\langle \xi_\text{L}(t) v_\text{c} (t) \big\rangle 
- \frac{\lambda_1}{\Delta \lambda}\left(\int_0^t  e^{\lambda_1(t-s)}\big\langle\xi_\text{L}(t)\xi_\text{L}(s)\big\rangle\,ds 
+ \int_0^t  e^{\lambda_1(t-s)}\big\langle\xi_\text{L}(t)\xi_\text{R}(s)\big\rangle\,ds\right)\\
\nonumber&\quad + \frac{\lambda_2}{\Delta \lambda}\left(\int_0^t  e^{\lambda_2(t-s)}\big\langle\xi_\text{L}(t)\xi_\text{L}(s)\big\rangle\,ds 
+ \int_0^t  e^{\lambda_2(t-s)}\big\langle\xi_\text{L}(t)\xi_\text{R}(s)\big\rangle\,ds\right)\\[1ex]
& = \frac{\gamma_\text{L} k_\text{B} T_\text{L}(t)}{m},\\
\label{eq:velnoisecorrhar}
\nonumber\big\langle \xi_\text{R}(t) v(t)\big\rangle &= 
\big\langle \xi_\text{R}(t) v_\text{c}(t) \big\rangle 
- \frac{\lambda_1}{\Delta \lambda}\left(\int_0^t  e^{\lambda_1(t-s)}\big\langle\xi_\text{R}(t)\xi_\text{L}(s)\big\rangle\,ds 
+ \int_0^t  e^{\lambda_1(t-s)}\big\langle\xi_\text{R}(t)\xi_\text{R}(s)\big\rangle\,ds\right)\\
\nonumber&\quad + \frac{\lambda_2}{\Delta \lambda}\left(\int_0^t  e^{\lambda_2(t-s)}\big\langle\xi_\text{R}(t)\xi_\text{L}(s)\big\rangle\,ds 
+ \int_0^t  e^{\lambda_2(t-s)}\big\langle\xi_\text{R}(t)\xi_\text{R}(s)\big\rangle\,ds\right)\\[1ex]
& = \frac{\gamma_\text{L} k_\text{B} T_\text{R}(t)}{m},
\end{align} 
\end{widetext}
here we have utilized $\big\langle \xi_\text{R}(t) v_\text{c}(t) \big\rangle = \big\langle \xi_\text{L}(t) v_\text{c}(t) \big\rangle = 0$ from causality by noting that all terms in $v_\text{c}(t)$ contain initial velocity $v_0$ or initial position $x_0$.

The expectation value for the energy of the system is
\begin{align}
\label{eq:energyHO}\big\langle E(t)\big\rangle  = \frac{1}{2} m \big\langle v^2(t)\big\rangle+ \frac{1}{2}m k \big\langle x^2(t)\big\rangle.
\end{align}
The second velocity moment $\big\langle v^2(t)\big\rangle$ can be evaluated by squaring the formal solution Eq.~(\ref{eq:vsolhar}) and applying the noise correlations leading to
\begin{widetext}
\begin{equation}
\begin{aligned}
\label{eq:vsqrHar}
\big\langle v^2(t)\big\rangle  &=  
 v^2_c(t)
 + \left(\frac{1}{\Delta \lambda}\right)^2\!\!\Bigg(\lambda^2_1\!\!\int_0^t\!\!\int_0^t  e^{\lambda_1(2t-s_1-s_2)}\big\langle\xi_\text{L}(s_1)\xi_\text{L}(s_2)\big\rangle\,ds_1\,ds_2 
 + \lambda^2_1\!\!\int_0^t\!\!\int_0^t  e^{\lambda_1(2t-s_1-s_2)}\big\langle\xi_\text{R}(s_1)\xi_\text{R}(s_2)\big\rangle\,ds_1\,ds_2 \\[1ex]
&  \quad + \lambda^2_2\int_0^t\!\!\int_0^t  e^{\lambda_2(2t-s_1-s_2)}\big\langle\xi_\text{L}(s_1)\xi_\text{L}(s_2)\big\rangle\,ds_1\,ds_2 
 + \lambda^2_2\int_0^t\!\!\int_0^t  e^{\lambda_2(2t-s_1-s_2)}\big\langle\xi_\text{R}(s_1)\xi_\text{R}(s_2)\big\rangle\,ds_1\,ds_2 \\[1ex] 
& \quad - 2 k\int_0^t\!\!\int_0^t   e^{\lambda_1(t-s_1)+ \lambda_2(t-s_2)}\big\langle\xi_\text{L}(s_1)\xi_\text{L}(s_2)\big\rangle\,ds_1\,ds_2
- 2 k\int_0^t\!\!\int_0^t   e^{\lambda_1(t-s_1)+ \lambda_2(t-s_2)}\big\langle\xi_\text{R}(s_1)\xi_\text{R}(s_2)\big\rangle\,ds_1\,ds_2 \Bigg)
\\[1ex]
&=  v^2_c(t)
- \left(\frac{1}{\Delta \lambda}\right)^2\!\!\Bigg(\frac{ \gamma k_\text{B} T}{m}\bigg(\lambda_1 \left(1-e^{2 \lambda_1 t}\right) +  \lambda_2\left(1-e^{2 \lambda_2 t}\right)
+ \frac{4 k(1- e^{- \gamma t})}{\gamma}\bigg)
\\[1ex]
& \quad  
+ \frac{2 k_\text{B} \lambda^2_1}{m}\bigg(\frac{ \gamma_\text{L} \Delta T_\text{L} ( 2 \lambda_1 \sin(\omega_\text{L} t) + \omega_\text{L} \cos(\omega_\text{L} t)- \omega_\text{L} e^{2 \lambda_1 t})}{4 \lambda_1^2+\omega^2_\text{L}} 
+ \frac{\gamma_\text{R} \Delta T_\text{R} (2 \lambda_1 \sin(\omega_\text{R} t) + \omega_\text{R} \cos(\omega_\text{R} t)- \omega_\text{R} e^{2 \lambda_1 t})}{4 \lambda^2_1+\omega^2_\text{R} }\bigg)\\[1ex]
& \quad 
+ \frac{2 k_\text{B} \lambda^2_2}{m}\bigg(\frac{ \gamma_\text{L} \Delta T_\text{L} (2 \lambda_2 \sin(\omega_\text{L} t) + \omega_\text{L} \cos(\omega_\text{L} t) - \omega_\text{L} e^{2 \lambda_2 t})}{4 \lambda_2^2+\omega^2_\text{L}} 
+ \frac{\gamma_\text{R} \Delta T_\text{R} (2 \lambda_2 \sin(\omega_\text{R} t) + \omega_\text{R} \cos(\omega_\text{R} t)-\omega_\text{R} e^{2 \lambda_2 t})}{4 \lambda^2_2+\omega^2_\text{R} }\bigg) \\[1ex]
& \quad
- k\frac{4 k_\text{B}}{m}\bigg(\frac{ \gamma_\text{L} \Delta T_\text{L} (\omega_\text{L} \cos(\omega_\text{L} t)-\gamma \sin(\omega_\text{L} t) - \omega_\text{L} e^{- \gamma t})}{\gamma^2+\omega^2_\text{L}}
+
\frac{ \gamma_\text{R} \Delta T_\text{R} (\omega_\text{R} \cos(\omega_\text{R} t) -\gamma \sin(\omega_\text{R} t)- \omega_\text{R} e^{-\gamma t})}{\gamma^2+\omega^2_\text{R}}\bigg)\Bigg),
\end{aligned}
\end{equation}
\end{widetext}
where we have excluded the cross-correlation terms between the two baths (which all are equal zero) in the top equation for brevity. 
The $\big\langle x^2(t)\big\rangle$ term in Eq.~(\ref{eq:energyHO}) is evaluated in a similar fashion to $\big\langle v^2(t)\big\rangle$ leading to:
\begin{widetext}
\begin{equation}
\begin{aligned}
\label{eq:xsqrHar}
\big\langle x^2(t)\big\rangle  &=  
 x^2_c(t)
- \left(\frac{1}{\Delta \lambda}\right)^2\!\!\Bigg(\frac{ \gamma k_\text{B} T}{m}\Bigg(\left(\frac{ 1-e^{2 \lambda_1 t}}{\lambda_1}\right) +  \left(\frac{1-e^{2 \lambda_2 t}}{\lambda_2}\right)
+ 4\left(\frac{1- e^{-\gamma t}}{\gamma}\right)\Bigg)
\\[1ex]
& \quad  
+ \frac{2 k_\text{B}}{m}\bigg(\frac{ \gamma_\text{L} \Delta T_\text{L} ( 2 \lambda_1 \sin(\omega_\text{L} t) + \omega_\text{L} \cos(\omega_\text{L} t)- \omega_\text{L} e^{2 \lambda_1 t})}{4 \lambda_1^2+\omega^2_\text{L}} 
+ \frac{\gamma_\text{R} \Delta T_\text{R} (2 \lambda_1 \sin(\omega_\text{R} t) + \omega_\text{R} \cos(\omega_\text{R} t)- \omega_\text{R} e^{2 \lambda_1 t})}{4 \lambda^2_1+\omega^2_\text{R} }\bigg)\\[1ex]
& \quad 
+ \frac{2 k_\text{B}}{m}\bigg(\frac{ \gamma_\text{L} \Delta T_\text{L} (2 \lambda_2 \sin(\omega_\text{L} t) + \omega_\text{L} \cos(\omega_\text{L} t) - \omega_\text{L} e^{2 \lambda_2 t})}{4 \lambda_2^2+\omega^2_\text{L}} 
+ \frac{\gamma_\text{R} \Delta T_\text{R} (2 \lambda_2 \sin(\omega_\text{R} t) + \omega_\text{R} \cos(\omega_\text{R} t)-\omega_\text{R} e^{2 \lambda_2 t})}{4 \lambda^2_2+\omega^2_\text{R} }\bigg) \\[1ex]
& \quad
- \frac{4 k_\text{B}}{m}\bigg(\frac{ \gamma_\text{L} \Delta T_\text{L} (\omega_\text{L} \cos(\omega_\text{L} t)-\gamma \sin(\omega_\text{L} t) - \omega_\text{L} e^{-\gamma t})}{\gamma^2+\omega^2_\text{L}} 
+
\frac{ \gamma_\text{R} \Delta T_\text{R} (\omega_\text{R} \cos(\omega_\text{R} t) - \gamma \sin(\omega_\text{R} t)- \omega_\text{R} e^{-\gamma t})}{\gamma^2+\omega^2_\text{R}}\bigg)\Bigg)
\end{aligned}
\end{equation}
\end{widetext}
The energy flux expressions are obtaining by substituting the correlation functions derived in this subsection into Eqs.~(\ref{eq:heatcurrentbathL})-(\ref{eq:heatcurrentbathS}).
We are mostly concerned with the long-time limit of these expressions in which the system approaches a TDNS. 
The energy flux expressions in the TDNS are:
\begin{widetext}
\begin{align}
\label{eq:JLlonghar}
\nonumber J_\text{L}(t) &= - \gamma_\text{L} k_\text{B} T_\text{L}(t) - \gamma_\text{L} \left(\frac{1}{\Delta \lambda}\right)^2\!\!\Bigg(k_\text{B} T \big(
4 k-\gamma^2\big)
\\[1ex]
\nonumber & \quad  
+ 2 k_\text{B} \lambda^2_1\bigg(\frac{ \gamma_\text{L} \Delta T_\text{L} ( 2 \lambda_1 \sin(\omega_\text{L} t) + \omega_\text{L} \cos(\omega_\text{L} t))}{4 \lambda_1^2+\omega^2_\text{L}} 
+ \frac{\gamma_\text{R} \Delta T_\text{R} (2 \lambda_1 \sin(\omega_\text{R} t) + \omega_\text{R} \cos(\omega_\text{R} t))}{4 \lambda^2_1+\omega^2_\text{R} }\bigg)\\[1ex]
\nonumber & \quad 
+ 2 k_\text{B} \lambda^2_2\bigg(\frac{ \gamma_\text{L} \Delta T_\text{L} (2 \lambda_2 \sin(\omega_\text{L} t) + \omega_\text{L} \cos(\omega_\text{L} t))}{4 \lambda_2^2+\omega^2_\text{L}} 
+ \frac{\gamma_\text{R} \Delta T_\text{R} (2 \lambda_2 \sin(\omega_\text{R} t) + \omega_\text{R} \cos(\omega_\text{R} t))}{4 \lambda^2_2+\omega^2_\text{R} }\bigg) \\[1ex]
& \quad
- 4 \lambda_1 \lambda_2 k_\text{B}\bigg(\frac{ \gamma_\text{L} \Delta T_\text{L} (\omega_\text{L} \cos(\omega_\text{L} t)-\gamma \sin(\omega_\text{L} t))}{\gamma^2+\omega^2_\text{L}} 
+
\frac{ \gamma_\text{R} \Delta T_\text{R} (\omega_\text{R} \cos(\omega_\text{R} t) -\gamma \sin(\omega_\text{R} t))}{\gamma^2+\omega^2_\text{R}}\bigg)\Bigg), \\[1ex]
\label{eq:JRlonghar}
\nonumber J_\text{R}(t) &= - \gamma_\text{R} k_\text{B} T_\text{R}(t) - \gamma_\text{R} \left(\frac{1}{\Delta \lambda}\right)^2\!\!\Bigg(k_\text{B} T \big(
4 k-\gamma^2\big)
\\[1ex]
\nonumber & \quad  
+ 2 k_\text{B} \lambda^2_1\bigg(\frac{ \gamma_\text{L} \Delta T_\text{L} ( 2 \lambda_1 \sin(\omega_\text{L} t) + \omega_\text{L} \cos(\omega_\text{L} t))}{4 \lambda_1^2+\omega^2_\text{L}} 
+ \frac{\gamma_\text{R} \Delta T_\text{R} (2 \lambda_1 \sin(\omega_\text{R} t) + \omega_\text{R} \cos(\omega_\text{R} t))}{4 \lambda^2_1+\omega^2_\text{R} }\bigg)\\[1ex]
\nonumber & \quad 
+ 2 k_\text{B} \lambda^2_2\bigg(\frac{ \gamma_\text{L} \Delta T_\text{L} (2 \lambda_2 \sin(\omega_\text{L} t) + \omega_\text{L} \cos(\omega_\text{L} t))}{4 \lambda_2^2+\omega^2_\text{L}} 
+ \frac{\gamma_\text{R} \Delta T_\text{R} (2 \lambda_2 \sin(\omega_\text{R} t) + \omega_\text{R} \cos(\omega_\text{R} t))}{4 \lambda^2_2+\omega^2_\text{R} }\bigg) \\[1ex]
& \quad
- 4 \lambda_1 \lambda_2 k_\text{B}\bigg(\frac{ \gamma_\text{L} \Delta T_\text{L} (\omega_\text{L} \cos(\omega_\text{L} t)-\gamma \sin(\omega_\text{L} t))}{\gamma^2+\omega^2_\text{L}} 
+
\frac{ \gamma_\text{R} \Delta T_\text{R} (\omega_\text{R} \cos(\omega_\text{R} t) - \gamma \sin(\omega_\text{R} t))}{\gamma^2+\omega^2_\text{R}}\bigg)\Bigg), \\[1ex]
\label{eq:JSlonghar}
\nonumber J_\text{sys}(t) &=  \left(\frac{1}{\Delta \lambda}\right)^2\!\!\Bigg(k_\text{B}\big(
\lambda_1^2+k\big)
\bigg(\frac{ \gamma_\text{L} \Delta T_\text{L} \omega_\text{L}(\omega_\text{L} \sin(\omega_\text{L} t) -2 \lambda_1 \cos(\omega_\text{L} t) )}{4 \lambda_1^2+\omega^2_\text{L}} 
+ \frac{\gamma_\text{R} \Delta T_\text{R} \omega_\text{R}( \omega_\text{R} \sin(\omega_\text{R} t)-2 \lambda_1 \cos(\omega_\text{R} t)) }{4 \lambda^2_1+\omega^2_\text{R} }\bigg)\\[1ex]
\nonumber & \quad + k_\text{B}\big(
\lambda_2^2+k\big)
\bigg(\frac{ \gamma_\text{L} \Delta T_\text{L}\omega_\text{L} (\omega_\text{L} \sin(\omega_\text{L} t) -2 \lambda_2 \cos(\omega_\text{L} t) )}{4 \lambda_2^2+\omega^2_\text{L}} 
+ \frac{\gamma_\text{R} \Delta T_\text{R} \omega_\text{R}( \omega_\text{R} \sin(\omega_\text{R} t)-2 \lambda_2 \cos(\omega_\text{R} t)) }{4 \lambda^2_2+\omega^2_\text{R} }\bigg)\\[1ex]
& \quad
- 4 \lambda_1 \lambda_2 k_\text{B}\bigg(\frac{ \gamma_\text{L} \Delta T_\text{L} \omega_\text{L} (\gamma \cos(\omega_\text{L} t)+\omega_\text{L} \sin(\omega_\text{L} t))}{\gamma^2+\omega^2_\text{L}} 
+
\frac{ \gamma_\text{R} \Delta T_\text{R} \omega_\text{R}(\gamma \cos(\omega_\text{R} t) + \omega_\text{R} \sin(\omega_\text{R} t))}{\gamma^2+\omega^2_\text{R}}\bigg)\Bigg).
\end{align}
\end{widetext}
Two observations are of note:
\begin{itemize}
\item Evaluating the net energy change over a period of temperature driving using Eq.~(\ref{eq:heat}), contributions from the trigonometric functions are zero. 
Therefore, for a single particle (a free particle or a particle in a harmonic potential) connecting two heat baths, the 
periodic temperature driving does not lead to enhanced energy transport in comparison to static temperature limit.
\item Due to the temperature oscillations, the energy fluxes depend on $k$. 
This differs from the single-particle case without temperature driving in which the energy fluxes are independent of $k$\cite{Lebowitz1959}.
\end{itemize}

In the quasistatic $(\omega_\text{L}/\gamma, \omega_\text{R}/\gamma) \to (0,0)$ limit, 
the derived analytical expressions reduce to the same energy flux expressions that were obtained for a free particle:
\begin{align}
\label{eq:JLlongsteadystatehar}
J^{(\text{QS})}_\text{L}(t) &=  k_\text{B} \frac{\gamma_\text{L} \gamma_\text{R}}{\gamma_\text{L} + \gamma_\text{R}} \Big(T_\text{R}(t)-T_\text{L}(t)\Big),\\[1ex]
\label{eq:JRlongsteadystatehar}
J^{(\text{QS})}_\text{R}(t) &= k_\text{B} \frac{\gamma_\text{L} \gamma_\text{R}}{\gamma_\text{L} + \gamma_\text{R}} \Big(T_\text{L}(t)-T_\text{R}(t)\Big), \\[1ex]
\label{eq:JSlongsteadystatehar}
J^{(\text{QS})}_\text{sys}(t) &= 0.
\end{align}
This implies that in the QS limit, the energy flux is independent of the potential form for a harmonic system.
In the static limit defined by $(\Delta T_\text{L}, \Delta T_\text{R}) \to (0,0)$, the system again reduces to the well-known form for the heat current of a single particle connecting two heat baths:
\begin{align}
\label{eq:JLlongsteadystate_notimehar}
J^{(\text{S})}_\text{L} &=  k_\text{B} \frac{\gamma_\text{L} \gamma_\text{R}}{\gamma_\text{L} + \gamma_\text{R}} \Big(T^{(0)}_\text{R}-T^{(0)}_\text{L}\Big),\\[1ex]
\label{eq:JRlongsteadystate_notimehar}
J^{(\text{S})}_\text{R} &= k_\text{B} \frac{\gamma_\text{L} \gamma_\text{R}}{\gamma_\text{L} + \gamma_\text{R}} \Big(T^{(0)}_\text{L}-T^{(0)}_\text{R}\Big),\\[1ex]
\label{eq:JSlongsteadystate_notimehar}
J^{(\text{S})}_\text{sys} &= 0,
\end{align}
which are independent of the potential form, 
as shown by Lebowitz \cite{Lebowitz1959}.

\begin{figure}[t]
\includegraphics[width = 8.5cm,clip]{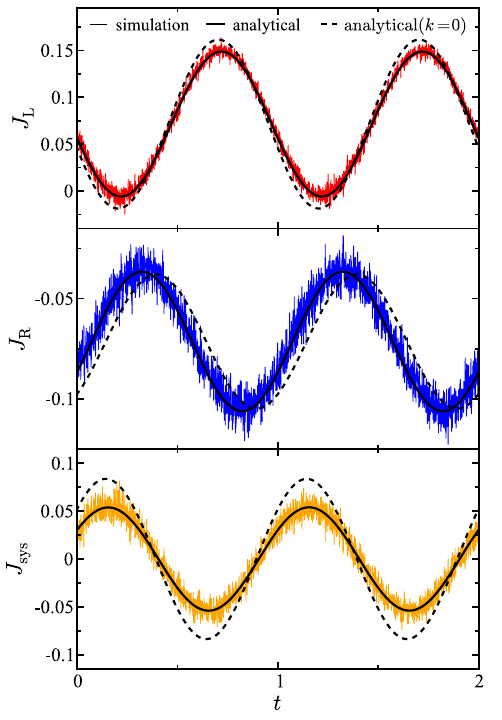}
\caption{\label{fig:Flux_1_har}
Time-dependence of the energy fluxes 
in the case of a particle in a harmonic potential connecting two heat baths with oscillating temperatures.
Each energy flux is shown in units of $\gamma k_\text{B} T$.
Time is shown in units of the total oscillation period $\mathcal{T}$.
Parameters are $k=5$, $\gamma = 2$ ($\gamma_\text{L} = 1$, $\gamma_\text{R} = 1$), $m = 1$, $T^{(0)}_\text{L} = 0.8$, $T^{(0)}_\text{R} = 1.07$,
$\Delta T_\text{L} = 0.2$, $\Delta T_\text{R} = 0$, $\omega_\text{L} = 5$, $\omega_\text{R} = 0$.
In each panel, the solid black curve is the exact analytical result, the dashed black curve is the analytical result in the free particle $k=0$ limit, and the noisy colored curves are the results generated from molecular dynamics simulations.
}
\end{figure}

Figure~\ref{fig:Flux_1_har} shows the time-dependent energy fluxes
for the left bath, right bath, and the system---in this case a particle moving in a harmonic potential.
As in the free-particle case,
the MD simulation results are in excellent agreement with the analytical results.
The dashed lines, representing the results for a free particle,
deviate from the results for a harmonic potential.
The difference between free particle and the harmonic particle 
are observed as changes in the energy flux oscillation phase 
and/or changes to the magnitude of oscillation.
For example, in the system energy flux, increasing $k$ away from the free particle $k = 0$ limit results in a reduced oscillation amplitude.

\begin{figure}[t]
\includegraphics[width = 8.5cm,clip]{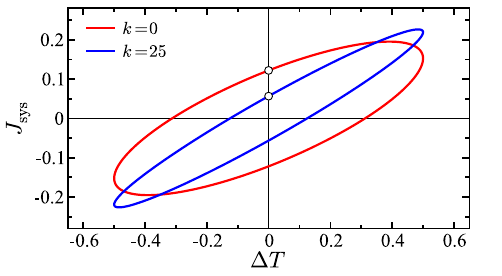}
\caption{\label{fig:Hys_1_har}
System energy flux as a function of temperature difference for a particle in a harmonic potential. 
Hysteresis loops are shown for $k=0$ (red wider loop) and $k = 25$ (blue more narrow loop).
The temperatures of the baths are $T^{(0)}_\text{R} =  T^{(0)}_\text{L} = 1$. 
Other parameters are the same as Fig.~\ref{fig:Flux_1}.
Each energy flux is shown in units of $\gamma k_\text{B} T$ and the temperature difference $\Delta T$ is shown in units of $T$.
The marker on each curve marks the value of the respective energy flux at $t=0$.
}
\end{figure}

The shape of the energy flux hysteresis loops are affected by the harmonic forces in the particle potential.
Figure~\ref{fig:Hys_1_har} shows hysteresis loops for two values of $k$ in the case
where one bath temperature is oscillating and the other is constant (the parameters are the same as in Fig.~\ref{fig:Hys_1}).
The characteristic shape of the loops are similar to the free-particle case
with the noticeable difference that the added external potential makes
the hysteresis loop narrower, a reflection of the reduced energy flux oscillation amplitude.

\begin{figure}[]
\includegraphics[width = 8.5cm,clip]{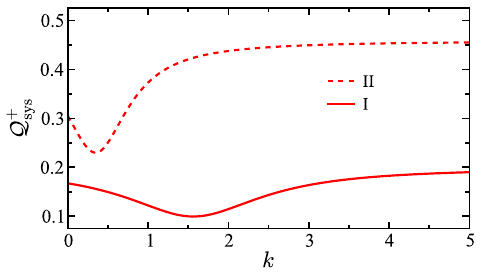}
\caption{\label{fig:Qplus_har}
Energy obtained by the system over one period of oscillation as function of $k$ for two sets of parameters denoted I and II.
Parameter set I, shown as a solid curve, are the same as in Fig.~\ref{fig:Flux_1}.
Parameter set II, shown as a dashed curve, are the same as in Fig.~\ref{fig:Flux_2}.
Energy is shown in units of $ k_\text{B} T$ and $k$ is shown in units of $\gamma^2$.
}
\end{figure}

The energy obtained by the system $\mathcal{Q}^+_\text{sys}$ 
over a period of oscillation is shown in Fig.~\ref{fig:Qplus_har}
as a function of the parameter $k$.
Results are shown for the  parameters used in Fig.~\ref{fig:Flux_1} (solid curve)
and in Fig.~\ref{fig:Flux_2} (dashed curve).
In both cases, the energy obtained by the system goes down first with increasing $k$,
then goes up approaching a horizontal asymptotic limit. 
The minima observed with respect to variation of $k$ values
are interesting features of the system.
No first-order parametric resonances are observed 
that correspond to the minima in the $\mathcal{Q}^+_\text{sys}$ vs. $k$ curves,  
meaning that the minima do not correspond to $k$ values equal to
$\omega^2_\text{L}$, $\omega^2_\text{R}$, $\gamma^2_\text{L}$, or $\gamma^2_\text{R}$.
Changing the friction parameters $\gamma_\text{L}$ and/or $\gamma_\text{R}$
alters the shape of the $\mathcal{Q}^+_\text{sys}$  vs. $k$ curve, but does not alter the location of the minimum.
However, the location of the minimum is dependent on the temperature oscillation frequencies. 
Generally, as the temperature oscillation frequencies are increased, the location of the minimum corresponds to a
larger value of $k$.
Possible higher-order resonance and off-resonance states 
may give rise to the dip in $\mathcal{Q}^+_\text{sys}$ with respect to variation of $k$.
The overall trends of the two parameter sets shown in Fig.~\ref{fig:Qplus_har} are similar,
but the case with two baths oscillating with different frequencies results in more energy being obtained by the system 
across the entire range of $k$. 
We conjecture this is due to constructive interference effects.

\section{Conclusions}
\label{sec:conclusion}
The energy transport properties of a particle connecting two heat baths 
with different temperatures where the temperature difference between baths is oscillating in time
have been derived.
A stochastic Langevin formalism has been applied to describe the
energy transport in different regimes of fast/slow temperature oscillations and
system-bath coupling strengths.
Analytical expressions have been derived for the time-dependent energy flux of each heat bath and for the system itself for two cases: (a) a free particle and (b) a particle in a harmonic potential. 
We find that the instantaneous energy fluxes (the heat currents) 
are time-dependent, as expected, due to the temperature oscillations 
and that in the long-time limit the system relaxes to a time-dependent nonequilibrium state.
Energy exchange between the system and the heat baths can be significantly 
affected by multiple factors such as the temperature oscillation frequency, the magnitude of the temperature oscillation,
and, in the case that the temperatures of both baths are oscillating, the ratio between the two oscillation frequencies.
In the limit of fast temperature oscillation relative to the relaxation rate of the system, 
significant differences are observed in the energy fluxes in comparison to the 
results obtained in the quasistatic limit defined by slow temperature oscillations.

The presented results illustrate that dynamical and inertial effects in the energy flux 
induced by fast temperature oscillations can give rise to complex
energy transport hysteresis effects. 
Our findings also suggest that applying time-periodic temperature modulations 
could be a possible pathway to control energy flow in molecular devices and nanoscale systems. 
The application of these effects in the design of thermal devices with memory is a potential future research direction.
In the case of a single particle, either a free particle or a particle in a harmonic potential, the thermal conductivity is not enhanced by the temperature oscillations in comparison to keeping the temperature gradient static. 
An increase in thermal conductivity due to temperature oscillations, which has been observed in macroscale systems, may arise at the molecular level due to anharmonicities in the system or interactions between multiples particles---both of which are not included in the present model but are targets for future work.

\section{Acknowledgements}
We acknowledge support from the Los Alamos National Laboratory (LANL) 
Directed Research and Development funds (LDRD).
This research was performed in part at the Center for Nonlinear Studies (CNLS) at LANL. 
The computing resources used to perform this
research were provided by the LANL Institutional Computing Program.

\bibliographystyle{apsrev}

\end{document}